\documentstyle[12pt,epsfig,multirow]{article}

\begin{document}

\begin{flushright}
\texttt{SINP/INO-2010/01}\\
\texttt{RKMVU/PHY-2010/01}\\
\end{flushright}
\bigskip

\begin{center}
{\Large \bf Interpreting the bounds on Solar Dark Matter induced
muons at Super-Kamiokande in the light of CDMS results}

\vspace{.5in}

{\bf Abhijit Bandyopadhyay$^\dagger$,
Sovan Chakraborty$^\ddagger$ and
Debasish Majumdar$^\ddagger$}\\
\vskip .5cm
{\normalsize \it $\dagger$ Ramakrishna Mission Vivekananda University,} \\
{\normalsize \it Belur Math, Howrah 711202, India}\\
\vskip 0.2cm
{\normalsize \it $\ddagger$ Saha Institute of Nuclear Physics,} \\
{\normalsize \it 1/AF Bidhannagar, Kolkata 700064, India}\\
\vskip 1cm
PACS numbers: 95.35.+d, 26.65.+t, 96.60.Jw 

\vskip 2cm

{\bf ABSTRACT}
\end{center}
We consider the recent limits on dark matter - nucleon elastic
scattering cross section from the analysis of CDMS II collaboration
using the two signal events observed in CDMS experiment. 
With these limits we try to 
interpret the Super-Kamiokande (SK) bounds on the detection rates
of up-going muons induced by the neutrinos 
that are produced in the sun from the decay of 
annihilation products of dark matter (WIMPs) captured 
in the solar core. 
Calculated rates of up-going muons for different annihilation channels 
at SK using CDMS bounds
are found to be orders below the predicted
upper limits of such up-going muon rates at SK.
Thus there exists room for enhancement (boost) of the calculated rates
using CDMS limits for interpreting SK bounds.
Such a  feature is expected to represent the PAMELA data 
with the current CDMS limits.
We also show the dependence of such a possible enhancement factor 
(boost) 
on WIMP mass for different
WIMP annihilation channels.
\newpage


Weakly interacting massive particles (WIMP) as dark matter in our 
galaxy can be trapped inside massive heavenly bodies like sun due to 
later's gravity \cite{use2}. This gravitational trapping may happen when such 
dark matter in course of its passage through sun undergoes elastic
scattering off the nuclei present in the solar core which causes its 
final velocity to fall below its velocity of escape from solar 
gravitational pull. These trapped dark matter may undergo the process
of pair-annihilation producing primarily $b$, $c$ and $t$ quarks, 
$\tau$ leptons, gauge bosons etc. Mass and composition of 
dark matter determine the annihilation products which in turn produce
 neutrinos and antineutrinos either through decay or pair annihilation.
Such neutrinos from the dark matter annihilation in solar environment 
have also been studied by previous authors (e.g. \cite{khlopov}).
Detection of such solar neutrinos in terrestrial detectors 
not only provides indirect evidence of dark matter but 
along with the results from the direct detection experiments of dark
matter such as CDMS \cite{cdms}, DAMA \cite{dama}, 
XENON \cite{xenon} etc. provides more insight into the nature 
of dark matter and its interactions. Analysis of recent observations
of two dark matter signal events with the Cryogenic Dark Matter 
Search experiment (CDMS II) at the Soudan Underground Laboratory 
 combined with all previous CDMS II data has been
reported by CDMS collaboration \cite{cdms1}. It sets new upper limits
on the WIMP-nucleon elastic scattering cross section 
($\sigma_\chi$) as a function of WIMP mass ($m_\chi$) 
\cite{cdms1}. The indirect searches for WIMPs through their annihilation 
in sun, with 1679.6 live days of data from SK detector using 
neutrino-induced upward through-going muons provide WIMP-induced 
upward muon flux limits at SK as a function of WIMP mass \cite{skwimp}. 
In this work we use the  90\% Confidence Level (C.L.) limits on 
$\sigma_\chi (m_\chi)$ from recent CDMS analysis \cite{cdms1} 
to calculate corresponding limits on detection rates of up-going
muons at SK as a function of  WIMP mass ($m_\chi$) and compare
them with the results in \cite{skwimp}.

%
The differential flux of neutrinos of type $i$ 
($i = \nu_\mu, \bar{\nu}_\mu$) at earth from  
WIMP annihilation products in the sun is given by \cite{jungman}
\begin{eqnarray}
{\left ( \frac {d\phi} {dE} \right ) }_i
&=& 
\frac {\Gamma_A} { 4\pi R^2} \sum_F B_F 
{ \left( \frac {dN} {dE} \right )}_{F,i}  
\end{eqnarray}
where $R$ is sun-earth distance and
$B_F$ is the annihilation branch for channel $F$.
${(dN/dE)}_{F,i}$ is the differential spectrum  of neutrinos
of type $i$ in the sun for the annihilation channel $F$.
The total rate for WIMP annihilation in the sun, 
$\Gamma_A$ is given in \cite{jasonkumar}
\begin{eqnarray}
\Gamma_A 
&=&
\frac{1}{2}C\tanh^2[{(aC)}^{1/2}\tau]
\end{eqnarray}
where $\tau \simeq 4.5$ Gyr is the age of sun.  
$a = \langle \sigma v \rangle / 4\sqrt{2} V$ is a function of
the  average WIMP annihilation cross section 
and the effective volume $V$ of WIMPs in the sun 
\cite{jasonkumar,use1,use2,use3,use4,use5}. Under
the astrophysical assumptions on density and velocity distribution 
as mentioned in \cite{jasonkumar,use1} (local dark matter density,
$\rho_\chi = 0.3 $ GeV cm$^{-3}$, mean velocity of dark matter, 
$\bar{v} = 300$ km sec$^{-1}$, 
a Maxwellian distribution of velocities etc.)
the dark matter capture rate $C$ in the sun is approximated as a 
function of the ratio of the WIMP-nucleus elastic scattering cross section
to the dark matter mass as \cite{jasonkumar,use2}
\begin{eqnarray}
C &\simeq& 10^{29} \frac{\sigma_\chi}{m_\chi}\ {\rm GeV}\ {\rm pb}^{-1} 
{\rm sec}^{-1}
\label{eqc} 
\end{eqnarray}
For such indirect detections of WIMPs at SK,
neutrinos  are detected through up-going muons produced by 
charged current interactions of neutrinos
with the rock below the detector. With numerical values of 
this cross section, the muon range in the rock and expressing
the energy distribution of neutrino flux in terms of its second 
moments, the total detection rate at SK detector of up-going muons 
induced by neutrinos from WIMP annihilation in the sun is given by 
\cite{jungman}
\begin{eqnarray}
\Gamma_{\rm detect} 
&=&
(1.27\times10^{-29}  {\rm yr}^{-1} ) 
\frac{C}{{\rm sec}^{-1}}
\left(\frac{m_\chi}{ {\rm GeV}}\right)^2
\displaystyle\sum_{i=\nu,\bar{\nu}} a_i b_i \sum_F B_F \langle {N_Z}^2 \rangle_{F,i} \times A_{eff}
\label{eqgammadetect}
\end{eqnarray}
where, $A_{eff} (\approx 1200 {\rm m}^2)$ 
is the muon effective area of the SK detector \cite{effarea},
$a_i$'s are the neutrino scattering coefficients.
The range of neutrino induced muons in the rock are given 
as the coefficients $b_i$. These coefficients are given by
$a_\nu = 6.8$, $a_{\bar{\nu}} = 3.1$, $b_\nu = 0.51$, 
$b_{\bar{\nu}} = 0.67$ \cite{jungman}. ${\langle {N_Z}^2 
\rangle}_{F,i}$'s are the second moments of the  spectrum
of neutrino type $i$ for the WIMP annihilation channel $F$ in the sun.
The  ${\langle {N_Z}^2 \rangle}_{F,i}$ for different channels relevant
for the present calculations are listed below \cite{jungman,ritz}.
\begin{enumerate}
\item[(a)] $\tau\bar{\tau}$ channel:
%
%
\begin{eqnarray}
{\langle {N_Z}^2 \rangle}_i (E_{\rm inj})
&=&
\Gamma_{\tau\to \mu\nu\bar{\nu}} h_{\tau,i}(E_{\rm inj}\tau_i) 
\quad (i = \nu_\mu,\bar{\nu}_\mu)
\end{eqnarray}
where $E_{\rm inj}$ is the injection energy of the decaying
WIMP annihilation product 
inside the sun, the branching ratio
$\Gamma_{\tau\to \mu\nu\bar{\nu}} \simeq 0.18$,
 $\tau_\nu (\tau_{\bar{\nu}}) = 1.01\times 10^{-3} 
 (3.8\times 10^{-4})$ GeV$^{-1}$ 
are the stopping coefficients for $\nu (\bar{\nu})$ and
\begin{eqnarray}
h_{\tau,\nu_\mu}(y) 
&=&
\frac{4+y}{30(1+y)^4} \nonumber\\
h_{\tau,\bar{\nu}_\mu}(y) 
&=&
\frac{168 + 354y + 348y^2 + 190y^3 + 56y^4 + 7y^5}{1260(1+y)} 
\end{eqnarray}
\item[(b)] $b\bar{b}$ channel:
%
%
\begin{eqnarray}
{\langle {N_Z}^2 \rangle}_i (E_{inj})
&\simeq &
\Gamma_{b\to \mu\nu X} \frac{\langle E_d \rangle^2}{E_i^2} 
h_{b,i}\left(\sqrt{\langle E_d^2 \rangle} \tau_i \right)
\end{eqnarray}
where, the  branching ratio $\Gamma_{b\to \mu\nu X} = 0.103$.
The hadronization and the decay processes of the quarks from
WIMP annihilation in the sun are characterized by the mean
energy $\langle E_d \rangle$ of the hadron, $\langle E_d \rangle =
E_c \exp\left(\frac{E_c}{E_0}\right)E_1\left(\frac{E_c}{E_0}\right)$,
$E_c = 470$ GeV \cite{ritz} . $E_0 = Z_f E_{\rm inj}$ is the initial hadron
energy for quarks injected with energy $E_{\rm inj}$ and  
$Z_f(=0.73)$ is the quenching fraction for $b$-quarks to account for
the loss of energy during hadronization, 
$E_1(x) = \int_x^\infty \frac{e^{-y}}{y}dy$. Also
$\langle E_d^2\rangle = E_c (E_0 - \langle E_d \rangle)$ and
$h_{b,i}$ is same as $h_{\tau,i}$.
%
%
%
%
%
%
%
\item[(c)] $W^+W^-$ and $Z\bar{Z}$ channel: 
%
%
\begin{eqnarray}
\left .{\langle {N_Z}^2 \rangle}_i (E_{\rm inj})\right|_{W}
&\simeq &
\left. \frac{\Gamma_{W\to \mu\nu}}{\beta} 
\frac{2+2E\tau_i(1+\alpha_i)+E^2\tau_i^2\alpha_i(1+\alpha_i)}
{E_{\rm inj}^3 \tau_i^3 \alpha_i (\alpha_i^2 - 1)(1+E\tau_i)^{\alpha_i+1}}
\right|^{E=E_{\rm inj}(1-\beta)/2}_{E=E_{\rm inj}(1+\beta)/2} \\
&& \nonumber\\
\left .{\langle {N_Z}^2 \rangle}_i (E_{\rm inj})\right|_{Z}
&\simeq &
\left. \frac{2\Gamma_{Z\to \nu_\mu\bar{\nu}_\mu}}{\beta} 
\frac{2+2E\tau_i(1+\alpha_i)+E^2\tau_i^2\alpha_i(1+\alpha_i)}
{E_{\rm inj}^3 \tau_i^3 \alpha_i (\alpha_i^2 - 1)(1+E\tau_i)^{\alpha_i+1}}
\right|^{E=E_{\rm inj}(1-\beta)/2}_{E=E_{\rm inj}(1+\beta)/2} 
\label{eqnzz}
\end{eqnarray}
where,
$\alpha_\nu(\alpha_{\bar{\nu}}) = 5.1(9.0)$ and the branching 
ratios $\Gamma_{W\to \mu\nu} = 0.105$ ,
$\Gamma_{Z\to \nu_\mu\bar{\nu}_\mu} = 0.067$. 
$\beta$ being the velocity of the gauge bosons.
\end{enumerate}

The injection energy $E_{\rm inj}$ is the
energy with which the WIMP annihilation products
$b\bar {b}$, $\tau\bar {\tau}$, $W$ and $Z$ etc. are produced.
In this work we present our results for two benchmark scenarios namely
$E_{\rm inj} = m_\chi$ and $\frac {m_\chi} {3}$.

Recently the CDMS collaboration announced two dark 
matter signal events with 90\% C.L.\cite{cdms1} 
which set new limits on dark matter-nucleon 
scattering cross-section $\sigma_\chi$ for
different $m_\chi$'s \cite{cdms1}. In 
this work we choose our set of $\sigma_\chi$'s and 
corresponding $m_\chi$'s from this limit
(exclusion plot for ``Soudan(all)" in Fig. 4 of Ref. \cite{cdms1}). 
With these CDMS limits on $\sigma_\chi$ we compute the detection 
rates for neutrino induced muons (from the product of the Dark 
Matter annihilation in the sun) for different annihilation channels.
Thus we calculate the CDMS induced limits on event rates in SK. 
We compute the detection rate for neutrino induced 
muons (from the product of WIMP annihilation
in the sun) using Eq. 4. 

\begin{figure}[t]
\includegraphics[width=8.0cm, height=7cm, angle=0]{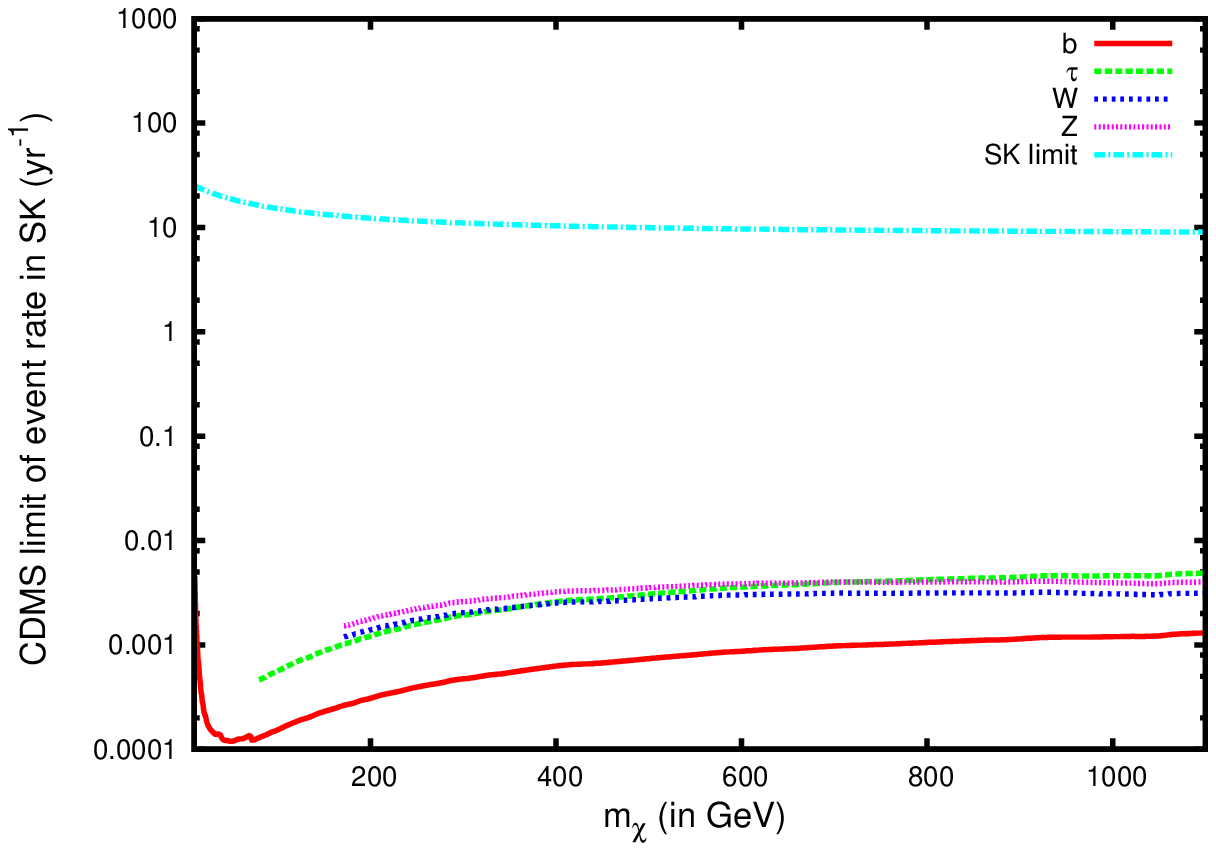}
\vglue -7.0cm \hglue 9.0cm
\includegraphics[width=8.0cm, height=7cm, angle=0]{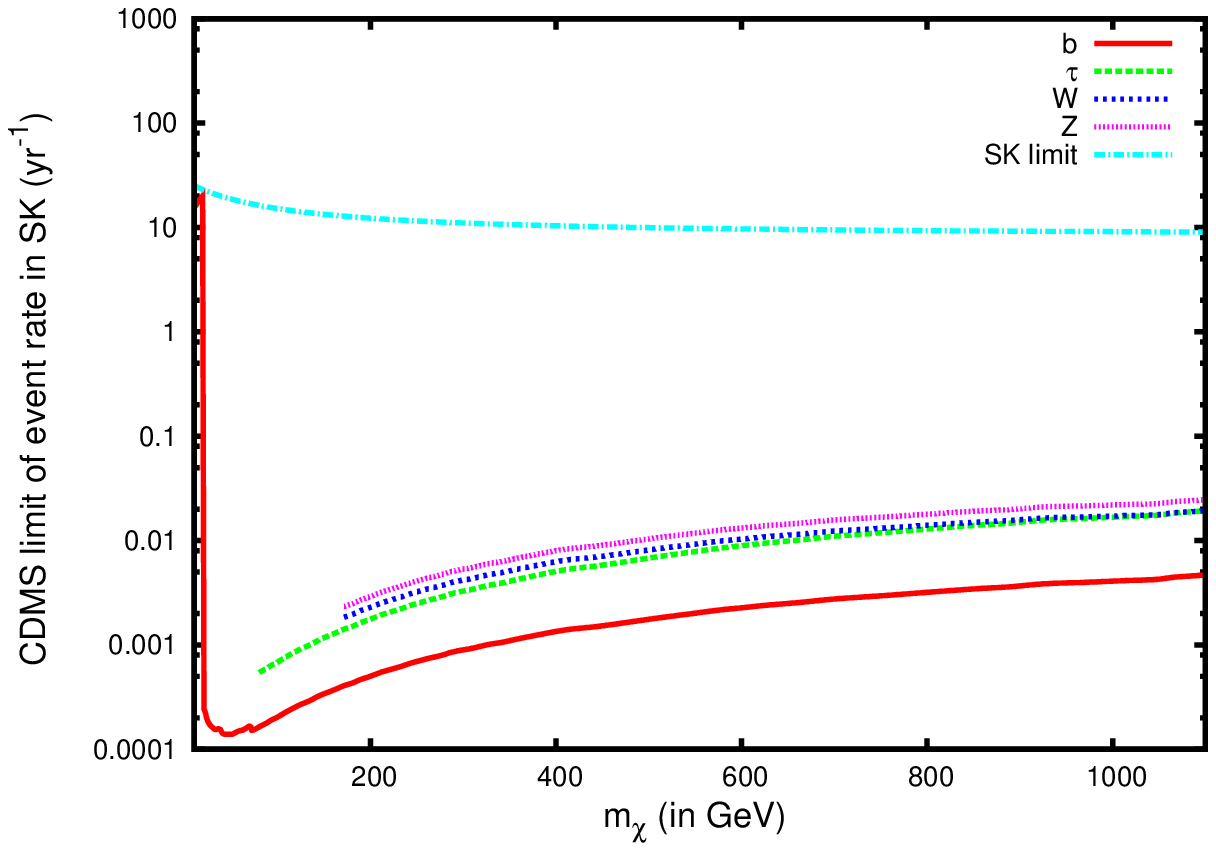}
\caption{\label{fig:rate}
CDMS upper bounds on WIMP induced muon event rates in different annihilation
channels with $E_{\rm inj} = m_\chi$ (left panel) and  
$E_{\rm inj} = m_\chi/3$
(right panel)}
\end{figure}
In this work we present our results for two benchmark scenarios namely
$E_{\rm inj} = m_\chi$ and $\frac {m_\chi} {3}$.
A conservative estimate for neutrino production
from WIMP annihilation in the sun is obtained by
considering the dominating effect of
$b{\bar{b}}$ production in the region $m_b < m_\chi < m_W$, that of
$\tau\bar{\tau}$ and $W, Z$  production in the regions
$m_W < m_\chi < m_t$ and
$m_\chi > m_t$ respectively \cite{jungman},
where $m_b$, $m_t$, $m_W$ are respective masses of
$b$-quark, $t$-quark and $W$-boson.
The  estimated limits of WIMP induced muon detection rates at SK using 
Eqs.\ (\ref{eqgammadetect}-\ref{eqnzz}) in the annihilation 
channels $b{\bar{b}}$ ,$\tau\bar{\tau}$ and $W$, $Z$
as a function of WIMP mass has been shown in
Fig.\ (\ref{fig:rate}). These rates are calculated 
for $E_{\rm inj} = m_\chi$
(left panel of Fig.\ \ref{fig:rate}) 
and $E_{\rm inj} = m_\chi/3$ (right panel of Fig.\ \ref{fig:rate}).
Plots for each channel in Fig.\ (\ref{fig:rate}) 
are obtained by taking the value of branching fraction 
($B_F$) for that channel to be $1$.
This can be justified by the fact that each channel is effective
in a particular $m_\chi$ region as discussed above. However, we have
shown in our results broader $m_\chi$ ranges for each of the channels. 
Also shown in this figure is the corresponding 
SK limit \cite{skwimp} and our results indicate existence for room
for an $m_\chi$ dependent enhancement of order 
$10^{3}-10^{5}$ of the detection rate 
in order to interpret SK bounds in terms of the CDMS 
$\sigma_\chi(m_\chi)$ limit. 
\begin{figure}[t]
\includegraphics[width=8.0cm, height=7cm, angle=0]{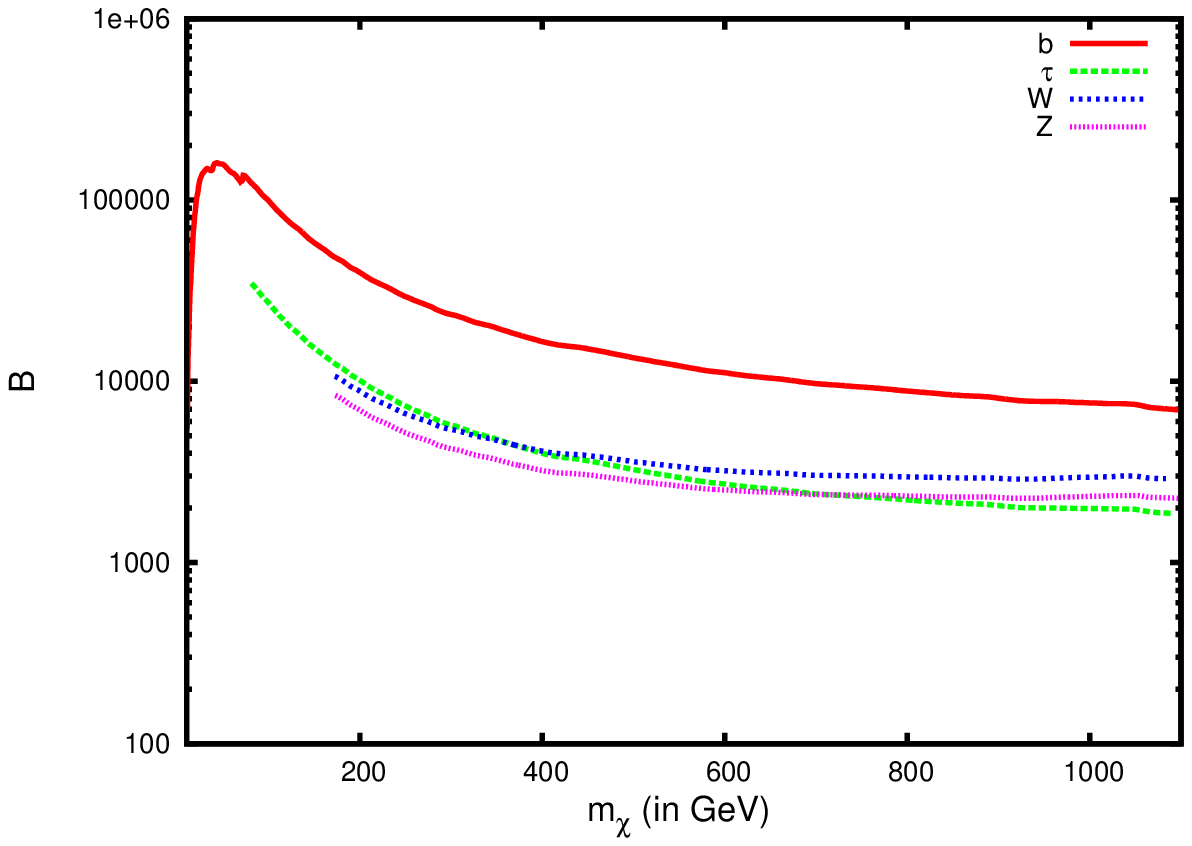}
\vglue -7.0cm \hglue 9.0cm
\includegraphics[width=8.0cm, height=7cm, angle=0]{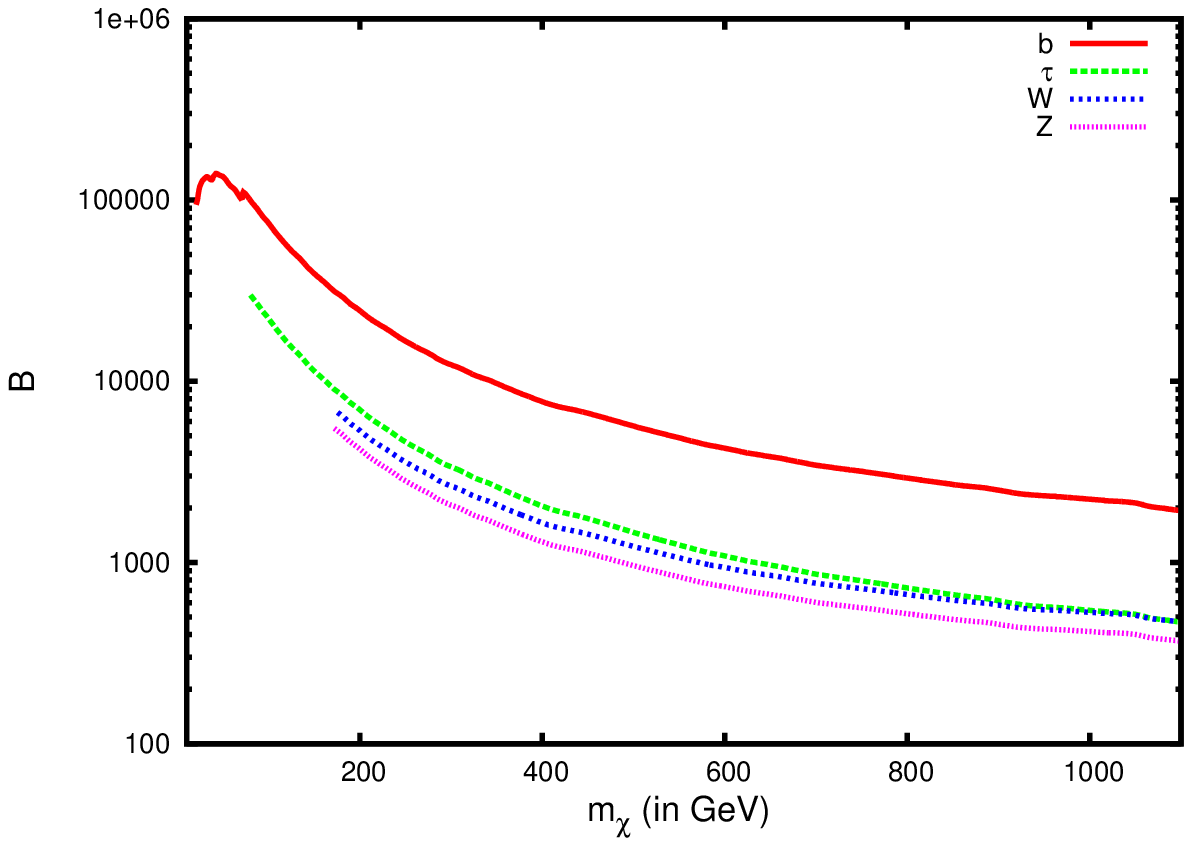}
\caption{\label{fig:b} Plot of $B$  vs $m_\chi$ in each
channel with $E_{\rm inj} = m_\chi$ (left panel) and  $E_{\rm inj} = m_\chi/3$
(right panel)}
\end{figure}

In a recent work, Cao et al \cite{caopamela} have shown that to 
study the anti-proton fraction of the results of 
satellite borne PAMELA experiment \cite{PAMELAnature, PAMELAprl} 
using the recent CDMS results
one needs to invoke a boost factor ($\sim 10^3$) 
for the annihilation of WIMPs into
quarks. This, in general is consistent with the enhancement needed to 
explain the excess of positron fraction as observed by PAMELA 
experiment \cite{PAMELAnature}. Our results also
indicate similar trends that if one uses the CDMS limit as is done in
the work, one is allowed to enhance the detection rate for the present
of SK bounds. We redefine $B_F$ in our framework (Eq. 4) as
$B \times B_F'$ where $B_F'$ is the branching fraction for a
particular process and $B$ is some boost factor. For a single channel
to represent the entire neutrino production process, ${B_F}^\prime = 1$.
Then we
estimate the upper limits of $B$'s for different dark matter annihilation
channels in the sun for the recent $\sigma_\chi - m_\chi$ CDMS limit. We do
this by comparing the detection rates of neutrinos -- 
obtained from the decays of the dark matter annihilation
products in the sun -- at Super-Kamiokande (SK) detector.
The upper limits on WIMP induced up-going muons at SK
as a function of WIMP mass is given by the SK Collaboration
\cite{skwimp}. 
In Fig.\ \ref{fig:b} (left panel) 
we show the variation of $B$ with $m_\chi$ for
$E_{\rm inj} = m_\chi$. Right panel of 
Fig.\ \ref{fig:b} shows similar plots for
$E_{\rm inj} = m_\chi$/3. 
From  Fig.\ \ref{fig:b} (left panel) we see $B$ varies
(with $m_\chi$) between $\sim 10^5 - \sim 10^4$  for $b\bar{b}$ channel,
$\sim 10^4 - \sim 10^3$ for $\tau\tau$ channel, $\sim 10^3$ for each of W and Z
channels etc.  For $E_{\rm inj} = m_\chi/3$ case (right panel of
Fig. \ref{fig:b})  
however, the variation of $B$'s for each of W and Z channels 
is between $\sim 10^3 - \sim 10^2$
and for $\tau\tau$ channel it is from $\sim 10^4 - 10^2$. 
These limits are consistent with the order $10^3$ as discussed earlier.\\
\begin{figure}[t]
\includegraphics[width=8.0cm, height=7cm, angle=0]{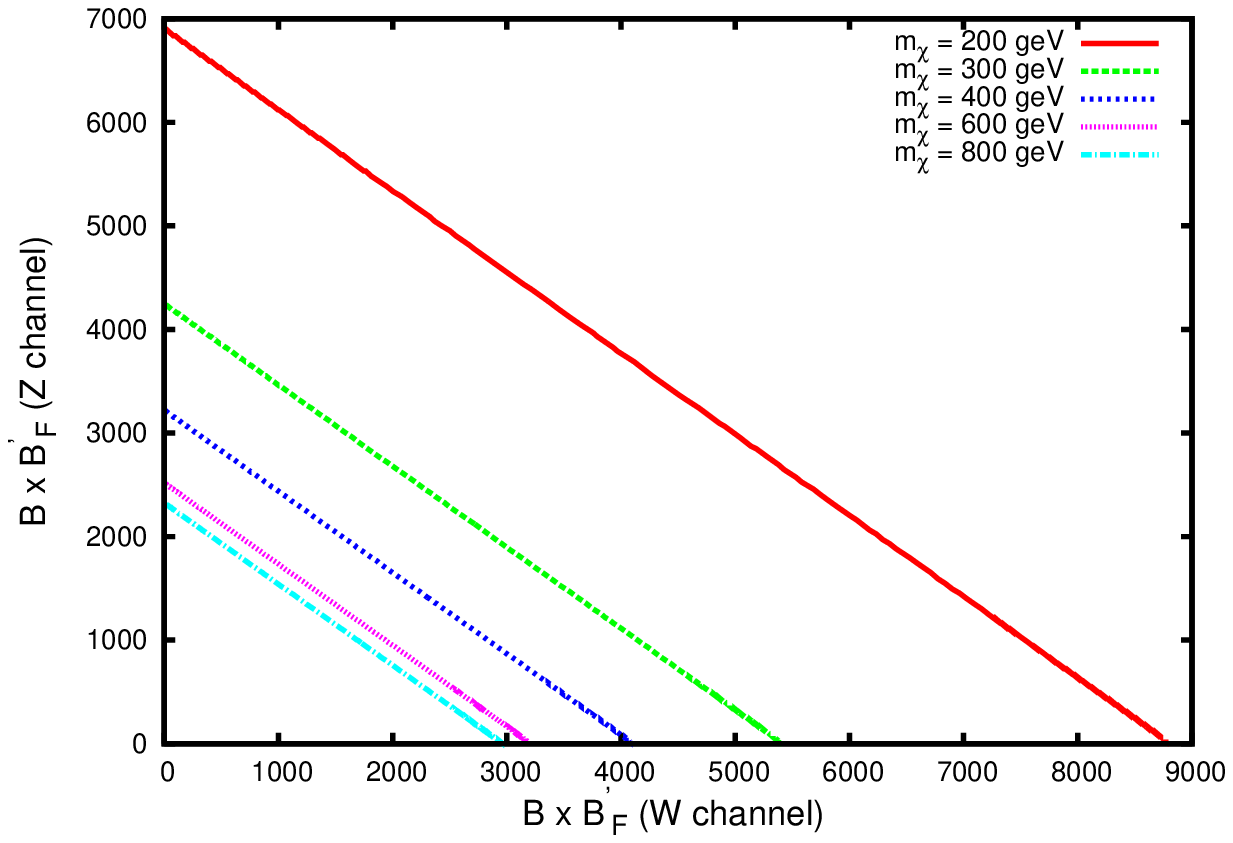}
\vglue -7.0cm \hglue 9.0cm
\includegraphics[width=8.0cm, height=7cm, angle=0]{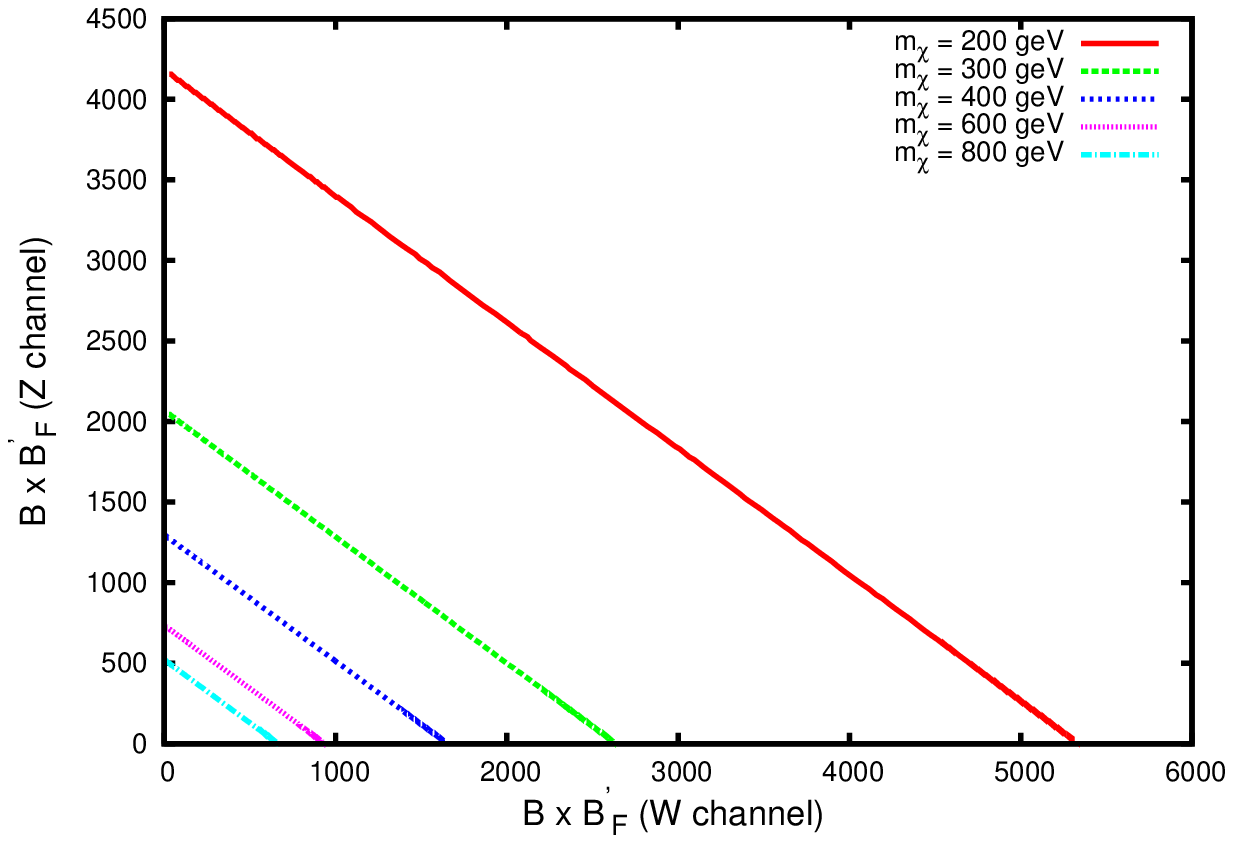}
\caption{\label{fig:wz} Iso $m_\chi$ plot in
$B\times {B_F}^\prime$ ($W$ channel)
-- $B\times {B_F}^\prime$ ($Z$ channel) plane interpreting SK limits
with $E_{\rm inj} = m_\chi$  (left panel) and  $E_{\rm inj} = m_\chi/3$
(right panel).}
\end{figure}
Although in  Figs.\ (\ref{fig:rate}-\ref{fig:b}) we have shown the
results for $W$ and $Z$ channels separately but as discussed earlier
both are dominant in the region $m_\chi > m_t$. Therefore 
we consider the effect of both of them together in calculating the rate
using Eq. (4).
In  Fig.\ \ref{fig:wz} (left panel) 
we present in $B\times {B_F}^\prime$ ($W$ channel)
-- $B\times {B_F}^\prime$ ($Z$ channel) plane the iso $m_\chi$ plots for
$E_{\rm inj} = m_\chi$ that represent the required rate to interpret the 
SK limits \cite{skwimp}. Right panel of 
Fig. \ref{fig:wz} gives the similar plots but with
$E_{\rm inj} = m_\chi/3$.   

In the present work we compute the WIMP induced upward going muon
rates at SK. They are induced by neutrinos from WIMP annihilation
products in the sun. We use recent CDMS bounds on WIMP-nucleon
scattering cross sections for different WIMP masses in our 
rate calculation. It is observed that representation of SK 
upper bounds on WIMP induced up-going muon rates allows an
enhancement in the calculated rates in all individual channels.
We made an estimation of this enhancement as a function of WIMP mass 
assuming branching fractions for each different channels to be 1 (maximum). 
Interpretation of PAMELA data in terms of CDMS limits 
also demand such enhancements of WIMP annihilation.
   
\vskip 5mm

The authors like to acknowledge the hospitality and facilities at 
XI Workshop on High Energy Physics Phenomenology (WHEPP XI)
at Physical Research Laboratory, Ahmedabad, India during which this work 
was initiated and progressed. 
The authors also thank S. Kundu and A. Ghosal for some useful discussions.


\end{document}